# Characterization and Optical Properties of Erbium doped $As_2S_3$ Films Prepared by Multi-layer Magnetron Sputtering

Wee Chong Tan, William T. Snider, Yifeng Zhou, Jaehyun Kim, Xiaomin Song, Travis James and Christi Madsen

Electrical and Computer Engineering Department,

Texas A&M University, 214 Zachry Engineering Center,

College Station, Texas 77843-3128, USA.

## Abstract

$As_2S_3$ film doped with erbium is prepared using multi-layer magnetron sputtering. The optical properties were measured by reflectance spectroscopy, and its chemical composition is examined by x-ray photoelectron, Rutherford backscattering, and Raman spectroscopy. The results show that the refractive index and absorption coefficient follow closely to a sputtered $As_2S_3$ film, and there are no detectable Er-S clusters and photo-induced $As_2O_3$ in the film. Rutherford backscattering spectroscopy shows that the film is homogeneous, and revealed the concentration level of erbium, and the stoichiometry of the film. The deposition method was used to fabricate an integrated Er-doped $As_2S_3$ Mach-Zehnder Interferometer and the presence of active erbium ions in the waveguide is evident from the green luminescence it emitted when it was pumped by 1488 nm diode laser. This method is attractive because the doping process can produce an Er:As$_2$S$_3$ film that is close to the ideal stoichiometry of As$_2$S$_3$ with lower risk of photo-decomposed As$_2$O$_3$ crystals developing on the surface when the as-deposited film is exposed to the environment.

## Introduction

Erbium (Er) is a chemical element in the lanthanide series that displays a very interesting property in its electronic configuration due to the shielding of the partially filled 4f shell by the electrons on the outer shells. This shielding has the effect that the energy levels of this 4f shell are largely insensitive to the environment that they reside in. Due to these unique optical properties, the elements in the lanthanide series are often employed in a wide variety of host matrices as an amplifying material for the field of photonics.

One of the most important uses of erbium is its inclusion into silica to form the erbium doped fiber amplifier (EDFA). The EDFA has become a very important component in optical telecommunications as they are routinely used for signal boosting during long-distance signal transmission to maintain signal quality. Chalcogenide hosts such as As$_2$S$_3$ is particularly suited as a host for erbium because this material exhibits very low phonon energies and does not inherently contain any hydroxyl or silicon oxide groups. Hosts with low-phonon energy hosts are desired for rare-earth ions because multi-phonon relaxation pathways can result in a rapid depopulation of the upper excited state and cause quenching of the photoluminescence.

There are a several ways to incorporate rare earth elements into chalcogenide films. It can be done by RF sputtering from custom made target or vacuum co-evaporation of chalcogenide glass and rare earth doped chalcogenide powder [1-3]. The custom-made sputtering target can either be from a commercial undoped target with erbium pieces on the surface, or made from grounding a rare earth doped chalcogenide bulk glass ingot into powder and hot pressing the power into a disk [1,2]. Similar methods have also been adopted to incorporate erbium into $As_2S_3$ thin film. To date, there are the co-evaporation method and the ion implantation method, which the $As_2S_3$ films were formed by thermal evaporation and the erbium doping was obtained by subsequent ion implantation [4,5]. However, an Er-doped $As_2S_3$ (Er:$As_2S_3$) film prepared by multi-layer magnetron sputtering is far more superior for use in making integrated optics in the following ways.

Due to the fact that we can independently control the sputtering rate of erbium directly in the deposition process, we believe an Er:$As_2S_3$ film prepared by multi-layer magnetron sputtering provides a more precise control of the doping concentration of erbium in the film. This is because deposition by sputtering is often very stable and reproducible, and so the doping rate of erbium will be very controllable too. Once calibrated, the concentration level can either be determined by the number of erbium layers in the film or by the sputtering rate. Moreover, the region with the highest concentration of erbium in the $Er:As_2S_3$ can be tailored to match the mode profiles of both the pump laser and the signal propagating in the waveguide. In this way, we can increase the absorption and emission coefficient of erbium in the $Er:As_2S_3$ and raise the efficiency of the waveguide amplifier.

Another major motivation for magnetron sputtering $Er:As_2S_3$ is that the process will produce an $As_2S_3$ film that is closer to their ideal stoichiometry. Any deviation from the ideal stoichiometry is undesired because it would not only change the optical properties of the film, e.g. refractive index, which is a critical parameter in waveguide design, but also made the material more susceptible to room temperature oxidation such as photo-induced $As_2O_3$ crystals. This is especially important for the fabrication of $As_2S_3$ waveguides, as very often the as-deposited $As_2S_3$ film prepared by resistive thermal evaporation or pulsed laser deposition, will oxidize into $As_2O_3$ when it comes into contact with oxygen in the ambient environment [6-8]. Even though a number of advanced processing methods have been proposed to solve these problems, the proposed methods in the literature still require the breaking of vacuum after the film is deposited and so will not effectively stop the oxidation process in the $As_2S_3$ film [9-12].

**Experimental Techniques**

Multi-layer sputtering of Er and $As_2S_3$ on $LiNbO_3$ substrate can be done by staggered deposition of erbium from a 2-inch 99.99% pure erbium target (American Elements Inc., Los Angeles, CA 90024)) and $As_2S_3$ from a 2-inch AMTIR-6 target (Amorphous

Materials, Inc., Garland, TX 75042) in a sputtering system. However, due to the need to keep both plasmas lit throughout the process and to balance a relatively slower deposition rate of $As_2S_3$ (~ 0.06 nm/s) with the tiny amount of erbium needed (~ 0.5 – 2.0 at. % Er), we will have to slow down the deposition rate of erbium by modulating its shutter during the run. Lowering the DC power alone is not enough to slow down erbium deposition rate, as the process is usually fixed at the working pressure and flow rate of argon of $As_2S_3$ due to the very demanding deposition conditions of magnetron sputtering $As_2S_3$ [13]. By setting the on and off time of the shutter, we are then able to maintain the deposition rate of erbium at a fractional rate of $As_2S_3$. This allows us to eventually control the desired concentration level of erbium in the composite film. Fig. 1 illustrates the multi-layer film structure created by this process. From the figure, we can see that there are two type of multi-layer $Er:As_2S_3$ film. If the shutter of the $As_2S_3$ target is closed during the deposition of erbium, the process is a regular multi-layered deposition, otherwise it is a process we called semi-cosputtering. Keeping the deposition of $As_2S_3$ running during the sputtering of erbium will help to spread out erbium ions and reduce the amount of Er-S clusters in $Er:As_2S_3$. This will prevent excessive cooperative up-conversion in an erbium-doped waveguide amplifier. The DC and RF power is respectively 25 and 35 W and the working pressure and argon flow rate is kept at 2.0 mTorr and 35 sccm throughout the deposition. Thermal annealing is carried out at 130 °C in a vacuum oven. All films were deposited onto a 1 $cm^2$ Si substrate. Films with different number of layers of erbium were prepared and tested.

The optical properties and thickness were measured by reflectance spectroscopy, and its chemical composition is studied using x-ray photoelectron spectroscopy (XPS), Rutherford backscattering spectroscopy (RBS), and Raman spectroscopy. All the binding energies obtained from the high resolution XPS scan were corrected to reflect the correct binding energy of carbon, C 1s, at 284.8 eV. RBS analysis is carried out with a 2 MeV He beam incident along the sample normal direction, with backscattered He atoms detected by a solid state detector with an energy resolution of 20 keV. The detector is located 165 degrees away from the beam incident directions. The Raman spectrum is obtained from a CCD detector with a spectral range of 400 to 950 nm. The excitation is done with a 633 nm He Ne laser through an objective lens set at 50 times.

**Results and Discussion**

In Fig. 2, we have the as-deposited refractive index of an $As_2S_3$ and two $Er:As_2S_3$ films with 4 and 16 layers of erbium in it. It shows that the refractive index of the $Er:As_2S_3$ films follow closely the $As_2S_3$ film. Moreover, increasing the number of erbium layers 4 times to increase the homogeneity of the film does not result in any drastic changes in the refractive index. In fact, the index of refraction looks almost identical to the sputtered $As_2S_3$ film. This is also true for the absorption coefficient.

Since the surface of the film is most susceptible to photo-induced $As_2O_3$ crystals, XPS is used to examine the chemical composition of the film at the surface. Of all the films deposited, only the one with a single layer of erbium has arsenic atom (3d shell) with binding energy of an $As_2O_3$, which according to NIST ranges from 43.9 eV to 46.3 eV. Fig. 3 shows the high resolution x-ray photoelectron spectroscopy spectrum of the arsenic 3d shell in $Er:As_2S_3$ film on a 1 cm$^2$ silicon wafer in their as-deposited and annealed states. All the binding energies were corrected to reflect the correct binding energy of C 1s at 284.8 eV.

Raman spectroscopy was also used to examine the chemical composition of our film. The spectrum of $As_2S_3$ film and the Si substrate were included as a reference for identifying the various shifts in the $Er:As_2S_3$ films. In Fig. 4, we can see that the Raman spectrum of our film with 16 layers of erbium showed no distinctive signs of the polycrystalline $Er_2S_3$. It implies that the dopant has not agglomerated into Er-S clusters inside the $As_2S_3$ host matrix when it is sputtered. This result is important because the clustering of Er in a host material will enhance the cooperative upconversion of Er and is detrimental to an Er-doped waveguide amplifier. The characteristic Raman signature of a polycrystalline $Er_2S_3$ can be found in the literature [14].

In the RBS measurement, the signal of each element can be clearly resolved in its own channel. To analyze the data, the erbium, arsenic and sulfur channels in the experimental spectrum were simulated with a box shaped profile as shown in Fig. 5. The analysis obtained from the fitting indicates that the film is homogenously doped as each element present has a constant concentration throughout the thickness of the film. The stoichiometry of the film is $Er_{0.7}As_{39.1}S_{60.2}$, which corresponds to an erbium concentration level of 2.35 wt. %, and its density is $3.6 \times 10^{22}$ atoms per $cm^3$. The thickness obtained from the simulation, $475 \pm 5$ nm, agrees with the measured thickness from reflection spectroscopy, $460 \pm 20$ nm, and the average thickness from a surface profiler, $467 \pm 21$ nm. The film has no detectable oxygen. The RBS spectrum of our sputtered Er:As$_2$S$_3$ look very similar to a homogeneous Er:As$_2$S$_3$ photo-resist fabricated by co-evaporation [5]. The RBS results clearly showed that multi-layer magnetron sputtering is able to produce homogeneous Er:As$_2$S$_3$ film with controlled stoichiometry.

In Fig. 6 we shows the typical green luminescence from an integrated Er:As$_2$S$_3$ MZI waveguide prepared by multi-layer magnetron sputtering. The 1488 nm pump signals are butt coupled into a 7 μm Ti:LiNbO$_3$ waveguide from a single mode fiber and then transfer over to the Er:As$_2$S$_3$ MZI by side coupling. The integrated Er:As$_2$S$_3$ MZI waveguide is pumped in both direction with a forward pump, which is travelling in the same direction as the signal, and a backward pump, which is travelling in the opposite direction to the signal. The maximum pump power at the fiber ends obtained in this way is about 182 mW. The Er:As$_2$S$_3$ contains eight layers of erbium deposited by semi-cosputtering As$_2$S$_3$ during erbium deposition. The film is thermally annealed at 130 C for 24 hrs, its thickness is $490 \pm 20$ nm and the Er:As$_2$S$_3$ waveguide is 3.5 μm wide. The measured concentration of erbium in the film calculated from RBS is found to be 0.4 atomic percent (at. %). The side coupling is accomplished with a two-stage taper. The tip width of the first stage taper varies linearly from 1.0 μm to 1.6 μm and the taper length is about 1000 μm long. The tip width of the second stage taper varies linearly from 1.6 μm to 3.5 µm and the taper length is also about 1000 µm long. The MZI was designed using the commercial software, OptiBPM, and consists of two cosine S-bends with one a mirror image of the other. The S-bend is 1500 µm wide and 300 µm high, and the spacing between the S-bends is 1000 µm. The calculated bending radius of the S-Bend is 380 µm. Fig. 7 shows the schematic diagram of the integrated MZI.

**Conclusion**

We have demonstrated the feasibility of incorporating erbium into $As_2S_3$ film deposited using multi-layer magnetron sputtering. The RBS spectrum of the $Er:As_2S_3$ film with 16 layers of erbium confirms the film is homogeneous and Raman spectroscopy shows no significant amount of Er-S clusters in the sputtered film. XPS analysis also revealed no amount of $As_2O_3$ crystals on the surface. The deposition method was used to fabricate an $Er:As_2S_3$ MZI waveguide and the presence of active erbium ions in the waveguide is evident from the green luminescence it emitted when it was pumped by 1488 nm diode laser.

**Acknowledgement**

We thank Dr. Shao Lin and his collaborators at the department of Nuclear Engineering of Texas A&M University for measuring the RBS spectrum for our film, and also NSF and DARPA for their financial support.

**Figures Caption**

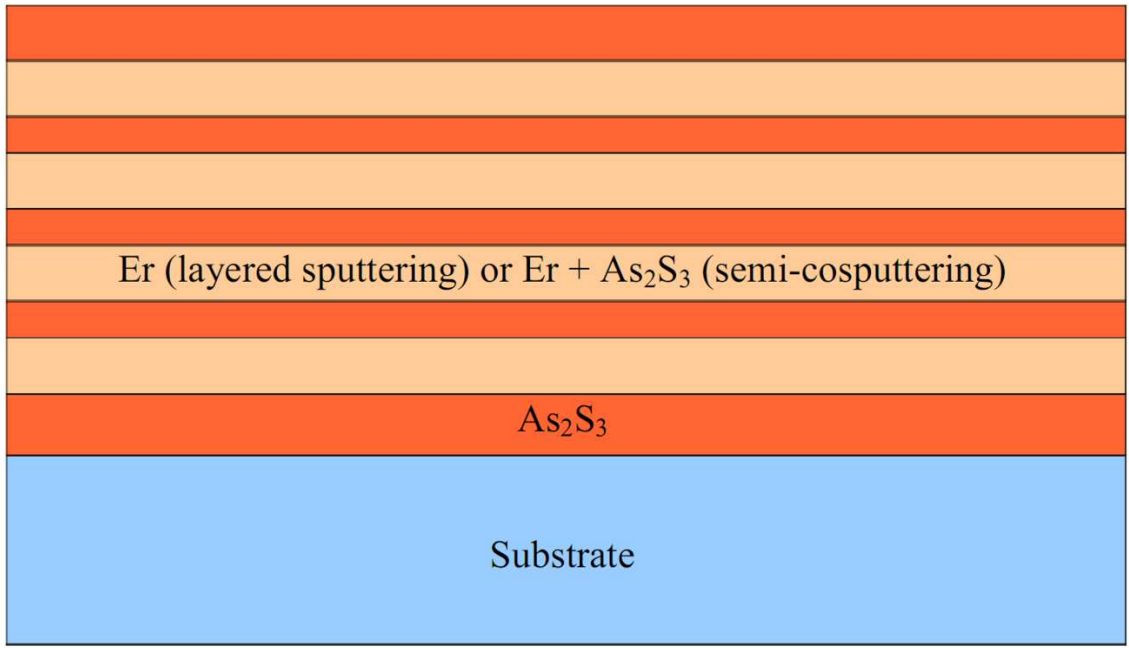

Fig. 1. The film structure of a multi-layer Er:As$_2$S$_3$ film prepared by staggered deposition. If the shutter of the As$_2$S$_3$ target is closed during the deposition of Er, the process is called layered sputtering, otherwise it is semi-cosputtering. Semi-cosputtering can help spread-out erbium and reduce Er-S clusters.

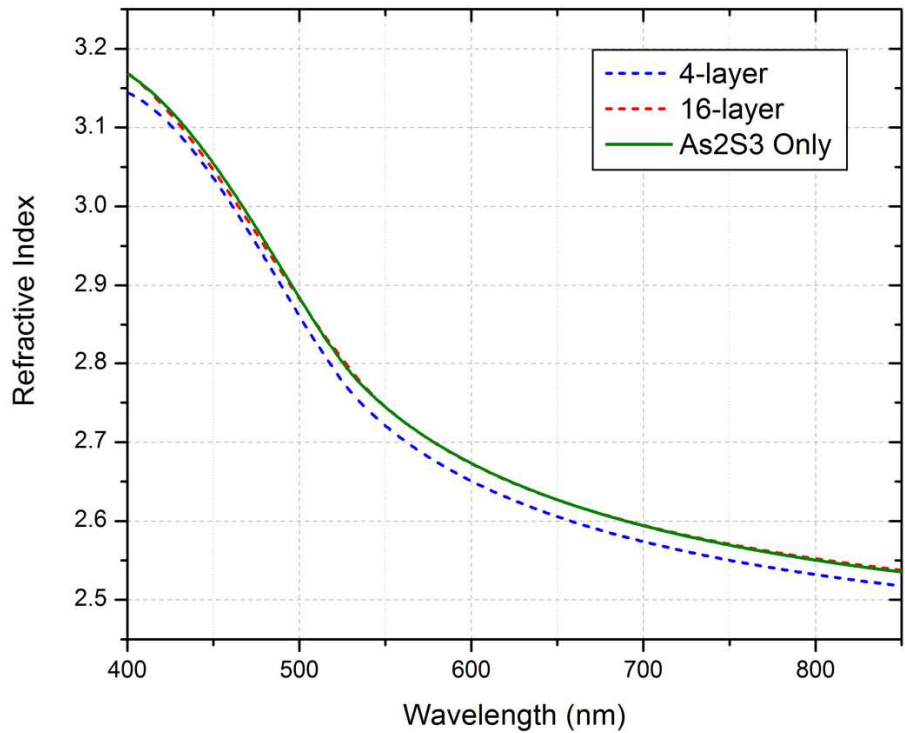

Fig. 2. The refractive index of as-deposited $As_2S_3$ and $Er:As_2S_3$ films prepared by magnetron sputtering. In the figure, As2S3 refers to magnetron sputtered $As_2S_3$ film. The thickness of the $As_2S_3$ film is $461 \pm 20$ nm, the $Er:As_2S_3$ with 4 layers of erbium (0.6 at. %) is $276 \pm 2$ nm and the $Er:As_2S_3$ with 16 layers of erbium (0.7 at. %) is $460 \pm 20$ nm.

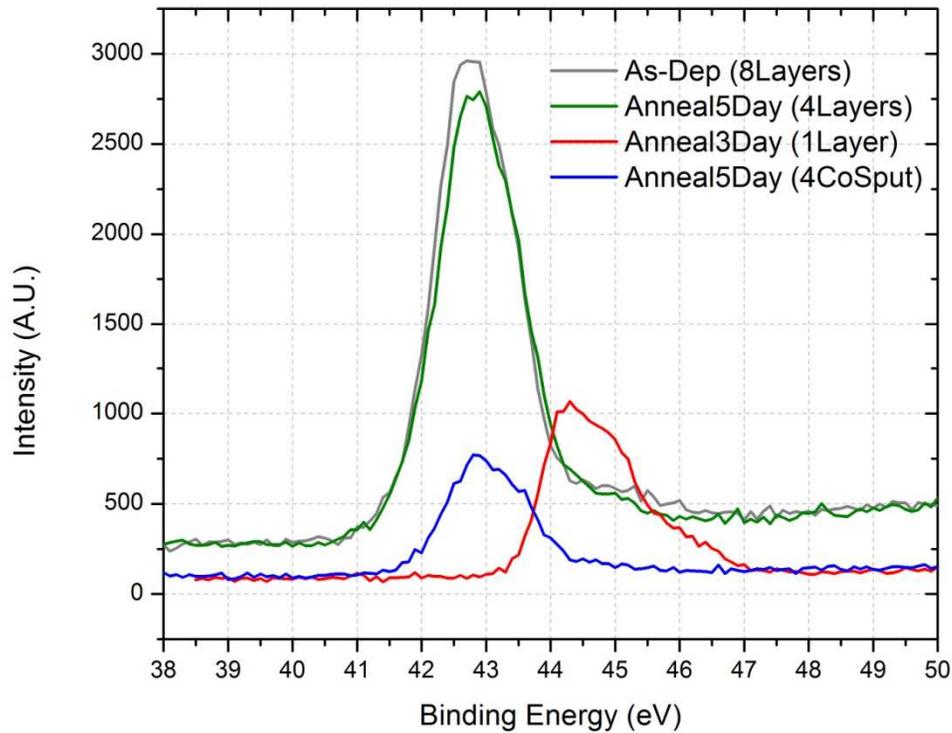

Fig. 3. High resolution XPS spectrum of As-3d shell in annealed Er:As$_2$S$_3$ films. In the figure, 4CoSput refers to film with four layers of semi-cosputtered Er. The thickness of the Er:As$_2$S$_3$ with one, four and eight layers of erbium are 456 ± 18 nm, 276 ± 2 nm and 523 ± 20 nm respectively. The film with four layers of semi-cosputtered erbium is 491 ± 20 nm.

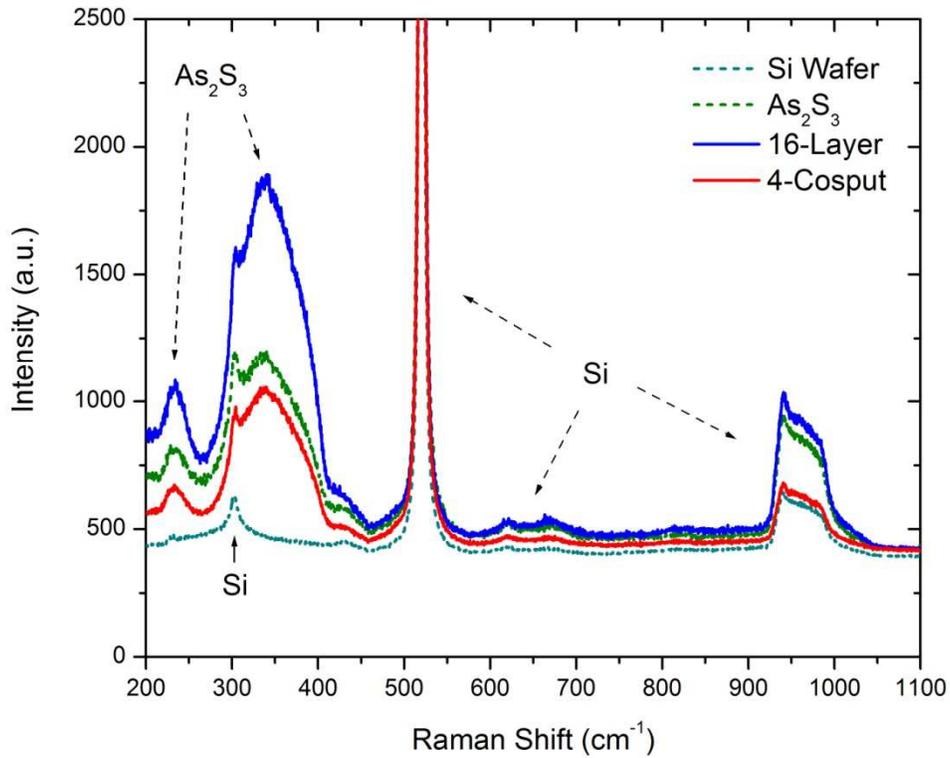

Fig. 4. Raman spectrum of as-deposited Er:As₂S₃ films on Si substrate. In the figure, 16-Layer refers to an Er:As₂S₃ films with sixteen layers of erbium and 4-Cosput refers to an Er:As₂S₃ films four layers of semi-cosputtered erbium. The thickness of the As₂S₃ film is $461 \pm 20$ nm. The film with sixteen layers of erbium is $460 \pm 20$ nm while the one with four layers of semi-cosputtered erbium is $491 \pm 20$ nm.

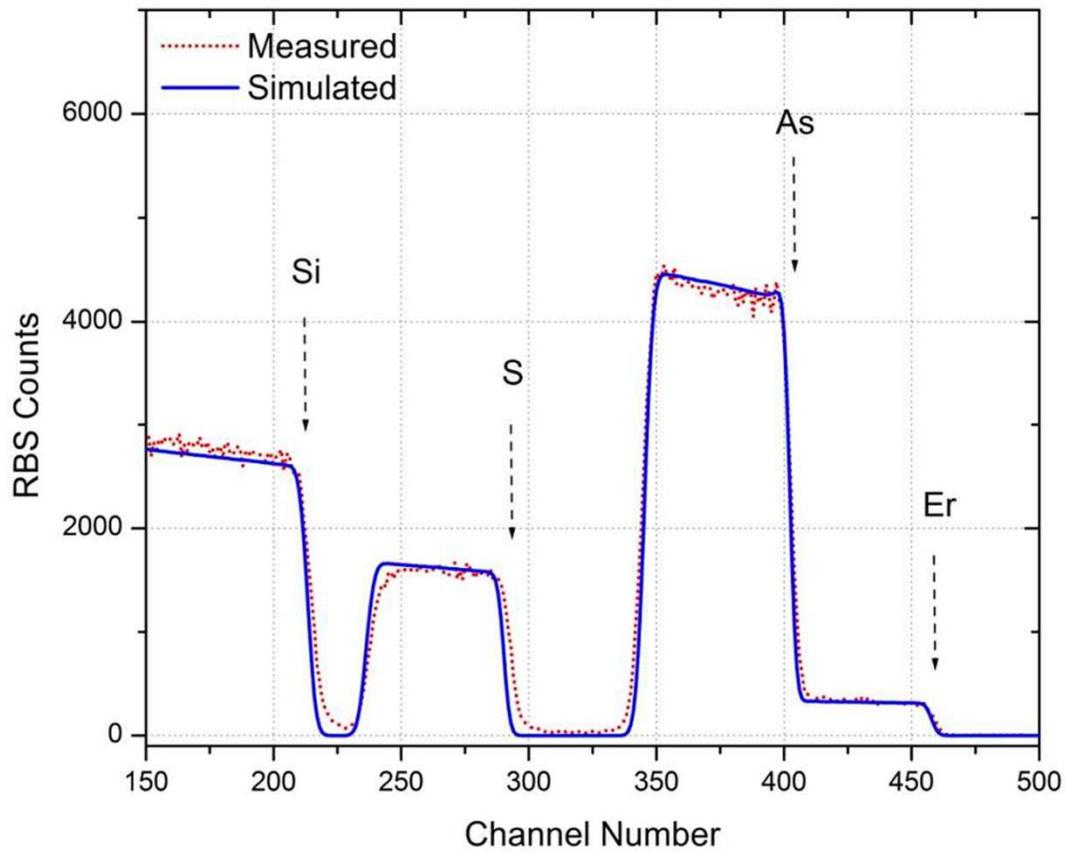

Fig. 5. The RBS spectra of a thermally annealed $Er_{0.7}As_{39.1}S_{60.2}$ film with 16 layers of erbium prepared by multi-layer magnetron sputtering. The film is homogeneous and the erbium concentration level is 2.35 wt. %

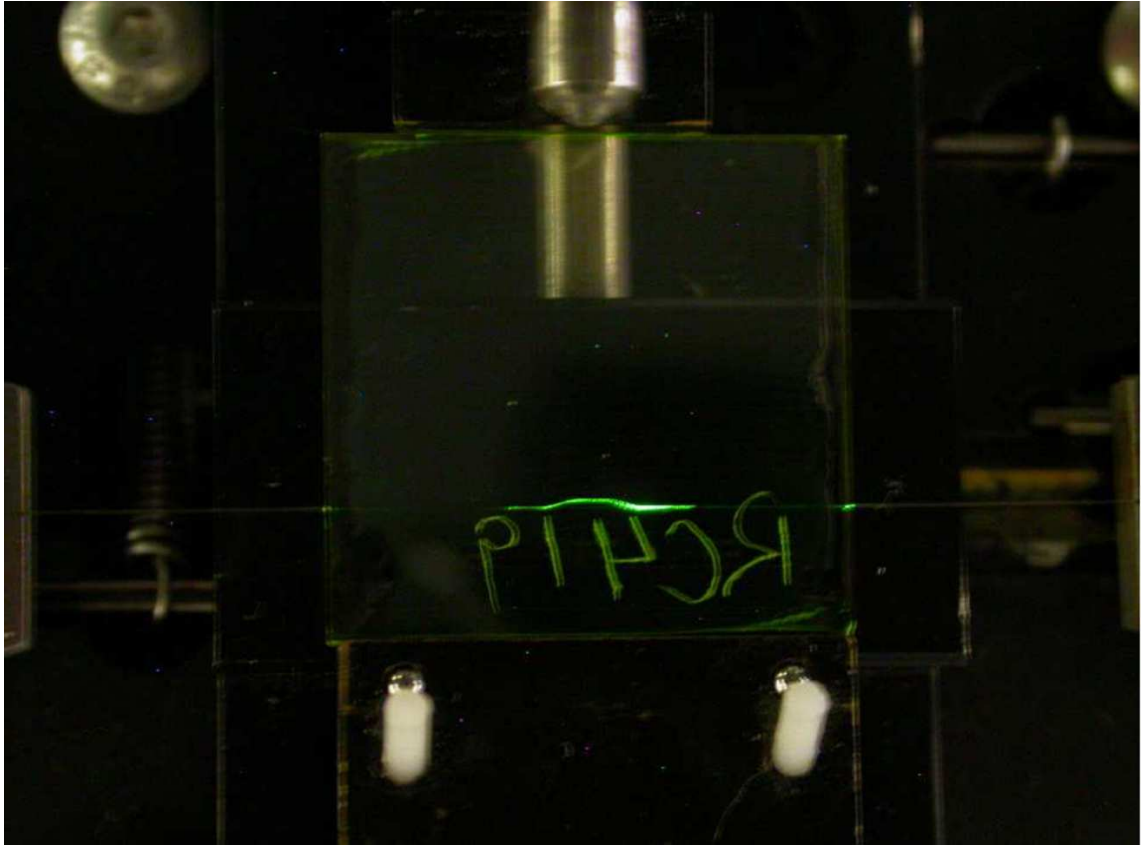

Fig. 6. The typical green luminescence from an Er:As$_2$S$_3$ MZI waveguide prepared by multi-layer magnetron sputtering. The Er:As$_2$S$_3$ contains 4 layers of cosputtered erbium. The film thickness is 490 ± 20 nm and the Er:As$_2$S$_3$ waveguide is 3.5 μm wide. The waveguide is thermally annealed for 24 hrs and Er concentration in the film is 0.4 at. %.

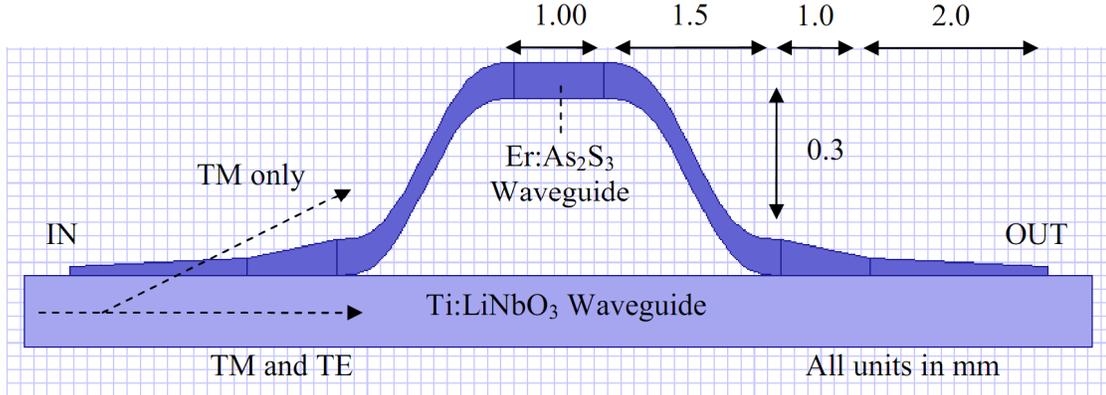

Fig. 7. The schematic drawing of an integrated Er:As$_2$S$_3$ MZI. A set of As$_2$S$_3$ S-Bend and reversed S-Bend is integrated with a straight Ti:LiNbO$_3$ to create a MZI. The width of the Ti:LiNbO$_3$ waveguide is 7 μm, the As$_2$S$_3$ waveguide is 3.5 μm, and the 2-stage As$_2$S$_3$ taper vary from 1-to-1.6 μm and then 1.6-to-3.5 μm. The arc length of the S-Bend is 1536.353 μm. TM refers to transverse magnetic propagation mode and TE refers to transverse electric mode.